\documentclass[sigconf]{acmart}

\AtBeginDocument{%
  }

\renewcommand\footnotetextcopyrightpermission[1]{}

\usepackage{graphicx}

\usepackage{microtype} 
\usepackage{amsmath}
\usepackage{array}
\usepackage{tabularx}
\usepackage{hyperref}
\usepackage{tikz}
\usetikzlibrary{arrows.meta,positioning}
\usetikzlibrary{fit}
\usepackage{listings}
\newcolumntype{L}[1]{>{\raggedright\arraybackslash}p{#1}}
\usepackage{tabularx,booktabs,siunitx,makecell,array}
\newcolumntype{Y}{>{\raggedright\arraybackslash}X}
\newcolumntype{R}{>{\raggedleft\arraybackslash}X}
\usepackage[T1]{fontenc}
\usepackage[utf8]{inputenc}
\usepackage{listings}

\usepackage{xcolor} 
\usepackage{orcidlink}

\usepackage{orcidlink}

\acmConference[e-Energy '26]
  {The 17th ACM International Conference on Future Energy Systems}
  {June 2026}
  {Banff, AB, Canada}

\acmBooktitle{The 17th ACM International Conference on Future Energy Systems (e-Energy '26), June 2026, Banff, AB, Canada}

\setcopyright{none}

\copyrightyear{2026}
\acmYear{2026}

\acmDOI{}
\acmISBN{}

\begin{document}

\title{Large Language Models as Explainable Cyberattack Detectors for Energy Industrial Control Systems}


\author{Weiyi Kong}
\affiliation{%
  \department{Department of Electrical and Computer Engineering}
  \institution{University of Toronto}
  \city{Toronto}
  \country{Canada}
}
\email{terry.kong@mail.utoronto.ca}

\author{Ahmad Mohammad Saber}
\affiliation{%
  \department{Department of Electrical and Computer Engineering}
  \institution{University of Toronto}
  \city{Toronto}
  \country{Canada}
}
\email{a.abdelsamie@mail.utoronto.ca}

\author{Amr Youssef}
\affiliation{%
  \department{Concordia Institute for Information Systems Engineering}
  \institution{Concordia University}
  \city{Montreal}
  \country{Canada}
}
\email{youssef@ciise.concordia.ca}

\author{Deepa Kundur}
\affiliation{%
  \department{Department of Electrical and Computer Engineering}
  \institution{University of Toronto}
  \city{Toronto}
  \country{Canada}
}
\email{dkundur@ece.utoronto.ca}


\begin{abstract}

In modern energy systems, industrial control systems (ICS) and power-system SCADA require intrusion detection that is not only accurate but also auditable by operators.
The ICS intrusion-detection landscape is currently dominated by established supervised detectors. 
In this paper, we study whether an off-the-shelf large language model (LLM) can serve as a complementary, human-in-the-loop layer for Modbus traffic.
We cast this as a binary network-side normal/critical decision task on two public ICS Modbus datasets, collapsing attack periods and other safety-critical behaviors into a single critical class. Each Modbus communication instance is converted into a compact token string derived from discretized protocol fields, and a prompt-configured LLM produces a normal/critical alert together with a concise, token-grounded incident record for analyst review.
Under matched event information and shared evaluation splits, the resulting LLM-based triage pipeline achieves high predictive performance on both benchmarks and is broadly comparable to strong supervised baselines, while requiring no task-specific weight updates.
To assess the audit record, we apply intervention-based diagnostics, including sufficiency- and necessity-style tests, which provide evidence that the cited tokens are often decision-relevant to the model's own prediction. These records are intended as audit signals rather than full human-grounded explanations.
\end{abstract}



\maketitle
\begin{center}
\small\textit{
Accepted to ACM EnergySP 2026, co-located with ACM e-Energy 2026.
This is the authors' accepted manuscript. The final published version
will appear in the ACM Digital Library.
}
\end{center}

\section{Introduction}

Industrial control systems (ICS) in modern power grids rely on SCADA protocols such as Modbus to exchange measurements and commands. These systems are vulnerable to cyberattacks including false-data injection (FDIA)~\cite{Liu2009FDIA}, and utilities need intrusion detection that is not only accurate but also auditable, so that alarms can be understood, trusted, and archived for incident response.

The ICS intrusion-detection literature is extensive, with supervised pipelines spanning a wide range of learners and Modbus-specific protocol models~\cite{Umer2022ICSSurvey,Hu2018ICSSurvey}. Supervised machine-learning detectors---from gradient-boosted trees to deep neural networks---already achieve strong accuracy on public ICS benchmarks~\cite{Umer2022ICSSurvey,Hu2018ICSSurvey}. Explainability approaches range from intrinsically interpretable detectors to post-hoc feature-attribution methods such as SHAP/LIME~\cite{Saber2025KAN_EV,Ahmad2024XAI_IDS_Industry5,Hosain2025XAIXGBoost}. In parallel, LLMs are increasingly explored for cybersecurity tasks including log triage and vulnerability analysis~\cite{Xu2025LLM4Security}, though most work uses the model as an assistant over natural-language reports rather than as a classifier over industrial protocol tokens.
Across these lines of work, a deployment-oriented gap remains. High-accuracy supervised detectors achieve strong predictive performance but, in many practical deployments, still involve site-specific labeled data collection, feature engineering, calibration, or retraining. Existing explainability methods can require model-specific diagnostics and do not always yield rationales in a form directly archivable as operator-facing incident records. And current LLM-for-security research has not systematically evaluated the LLM as a classifier over discretized ICS protocol tokens under matched inputs and controlled baselines.

This paper addresses this gap by studying an off-the-shelf LLM as a triage interface over a compact Modbus token string. We do not seek to replace established supervised approaches, nor do we position the LLM as an autonomous final decision-maker. Instead, we study whether a general-purpose, off-the-shelf LLM, driven only by a compact text prompt encoding generic Modbus semantics and ICS safety cues, can serve as a human-in-the-loop triage layer: flagging Modbus traffic as normal or critical so that operators can prioritize investigation, while the model simultaneously emits a concise, archivable incident record for each alert.
We formulate this as a binary normal/critical task on two public datasets---the LeMay CSET'16 Modbus dataset~\cite{LeMay2016SCADA} and CIC Modbus 2023~\cite{CICModbus2023}---by collapsing all attack periods and safety-critical behaviors into a single \emph{critical} class. This scope is deliberate: operators often need a conservative gate that surfaces traffic warranting review, while multi-class attack attribution remains the responsibility of downstream analysis by human analysts, specialized classifiers, or both.
On the data-efficiency side, the LLM requires no task-specific weight updates, and its behavior is configured through a prompt with a small number of prototypical examples (few-shot); we show that this prompt-only approach achieves detection performance broadly comparable to supervised baselines trained on the full task-specific training split. On the auditability side, a second-pass auditor produces a structured record containing verbatim evidence tokens, risk-category tags, and an optional counterfactual edit. We evaluate this record using intervention-based diagnostics---sufficiency, necessity, and counterfactual probes---that test whether cited tokens are decision-relevant to the classifier~\cite{DeYoung2020ERASER,Verma2024CFRecourseReview}. We note that this constitutes a token-grounded decision-relevance audit, not a fully human-grounded explanation; improving interpretability through domain-expert evaluation is an important direction for future work.

Our contributions are threefold. First, we evaluate the feasibility of a prompt-configured, few-shot LLM as a complementary triage layer for ICS Modbus intrusion detection on two public datasets, showing competitive performance with supervised baselines without any task-specific weight updates. Second, we design a two-pass pipeline separating prediction from explanation, producing a structured, operator-facing incident record grounded in verbatim protocol tokens. Third, we assess explanation quality using intervention-based diagnostics, providing quantitative evidence that cited tokens are decision-relevant. Across both benchmarks, the LLM attains approximately 0.98 accuracy with high recall on the critical class; the practical value lies in the combination of competitive detection, prompt-only deployment, and an auditable record suitable for alerting under human review. Section~\ref{sec:methodology} presents the methodology; Section~\ref{sec:dataset} describes datasets and evaluation; Section~\ref{sec:results} reports results; and Section~\ref{sec:conclusion} concludes the paper.

\section{Threat Model and Deployment Scope}
We consider a network-side adversary who can inject, replay, modify, or flood Modbus traffic, as well as send anomalous read/write requests. Real-world grid incidents have shown that such intrusions into power-system operations can have substantial operational consequences~\cite{CISA2021UkraineICS}. Our objective is therefore not full attack attribution or autonomous response, but conservative triage: identifying traffic that should be surfaced for prompt human review.
The proposed system is deployed as a passive decision-support layer over observed Modbus messages. It does not issue control commands, block traffic, or replace downstream forensic analysis. In this setting, false negatives may delay detection of malicious or safety-critical traffic, whereas false positives primarily increase analyst workload and alert fatigue. Privacy and data-governance considerations in API-based deployment are important but outside the scope of this offline evaluation on public datasets.

\section{LLM-Based Detection and Auditing Methodology}
\label{sec:methodology}

\subsection{Overview}
\begin{figure}[t!]
    \centering
    \includegraphics[width=\columnwidth]{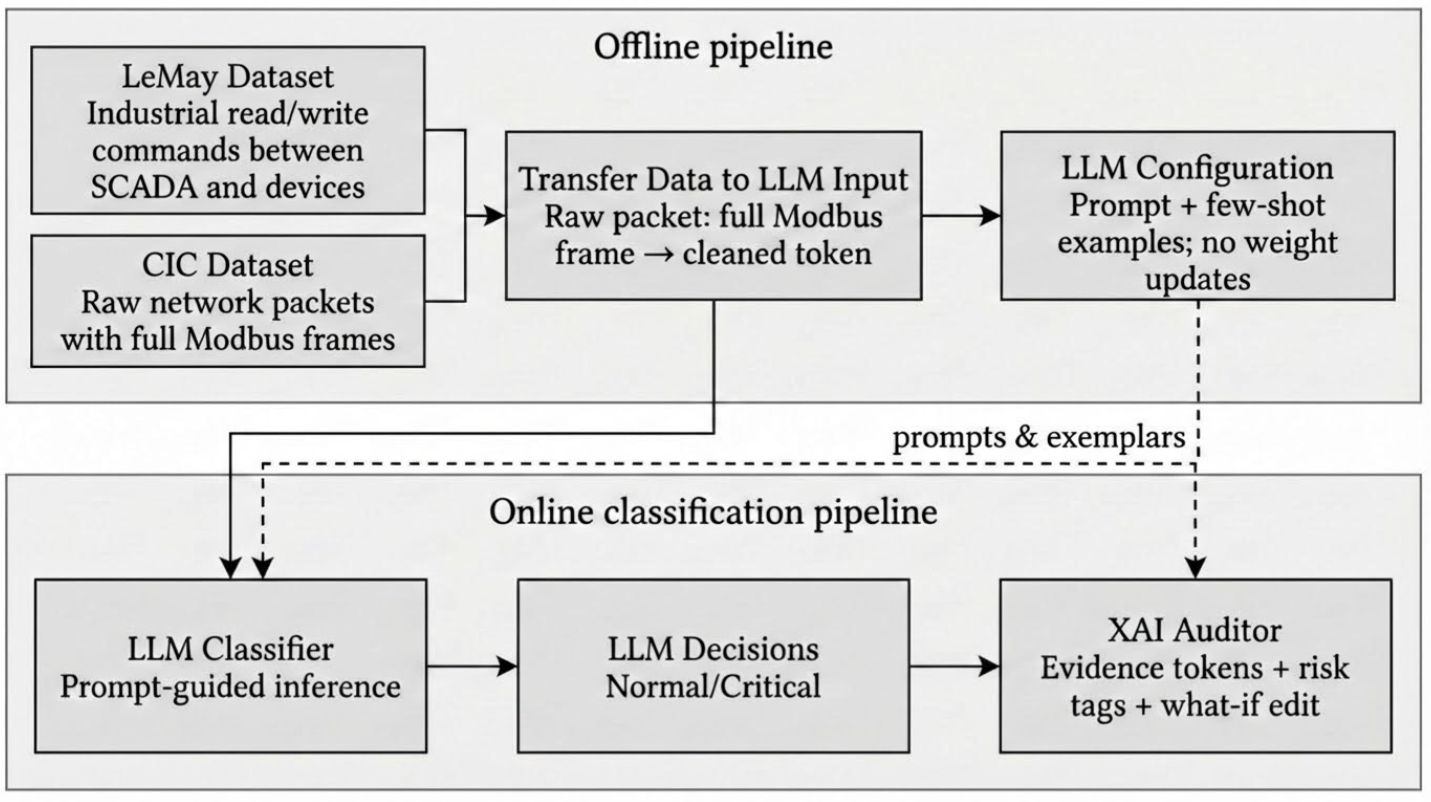} 
    \caption{Overview of the proposed framework. Offline preparation converts Modbus/TCP traces from LeMay and CIC into a unified token-string representation (no model training). Online inference uses a prompt-guided LLM to output a binary label (normal/critical), followed by an auditing step that generates supporting evidence tokens, risk tags, and optional what-if edits.}
    \label{fig:system_pipeline}
\end{figure}

An overview of the proposed framework is illustrated in Fig. \ref{fig:system_pipeline}.
We study intrusion detection on Modbus network traffic as a binary normal-versus-critical task using two public traces (LeMay and CIC). A shared ingestion pipeline normalizes timestamps and extracts discretized protocol and timing evidence, and all methods use the same canonical train/validation/test split. For the LLM-based approach,  it is guided by a text prompt that encodes generic Modbus semantics and high-level ICS safety cues to output a normal/critical decision together with a short, operator-readable record. Prediction and explanation are computed in two passes: the first pass outputs the binary label, and the second pass asks the model to produce a brief
explanation by copying a few key tokens and, when possible, suggesting a minimal “what-if” change to the tokens that would change the decision. Evaluation follows a unified protocol across methods, reporting accuracy, macro-F1 (unweighted average of per-class F1 scores), and recall/F1 on the critical class on the shared held-out test split.

\subsection{LLM and Problem setup}

We instantiate the proposed pipeline with GPT-4o~\cite{OpenAI2024GPT4oSystemCard}, used here as an off-the-shelf general-purpose LLM accessed via API in text-only mode. No task-specific fine-tuning, weight updates, or domain-specific retraining are performed; instead, all task behavior is specified through prompting, including a small number of in-context exemplars.
We use temperature-0 decoding to reduce output variability across repeated calls.

Given a compact text sequence $x$ representing a Modbus frame, the goal is to predict a binary label
$y \in \{\text{normal}, \text{critical}\}$ and to produce an auditable explanation tied to the input.
Here we use “critical” as a unified positive class: for each dataset, all traffic that occurs during any attack
scenario or other safety-critical behavior marked in the original logs is labeled critical, and all remaining
traffic is labeled normal (see Section~\ref{sec:dataset} for per-dataset details).
We use an off-the-shelf LLM without modifying its parameters; everything that differs between datasets is encoded in a compact domain prompt configuration that describes generic Modbus behavior and ICS safety cues. We refer to this collection of prompt text and cues as the prompt configuration, denoted by $\pi$.
The resulting classifier is written as

\begin{equation}\label{eq:clf}
f(x;\pi) = \bigl(\hat{y}(x),\, s(x)\bigr)
\end{equation}
Here $s(x)\in[0,1]$ is an auxiliary score recorded for analysis. For the LLM-based classifier,
$\hat{y}(x)\in\{\text{normal},\text{critical}\}$ is taken directly from the model's discrete output, and
$s(x)$ is the confidence value reported in the JSON response. We do not calibrate this confidence and do not
use it to choose the label; it is stored only for later analysis of explanation behavior.

For supervised baselines that output a scalar score $s_{\mathrm{base}}(x)$ for the critical class (for example,
the sigmoid probability in logistic regression), the predicted label is obtained by thresholding this score.
We sweep thresholds $\tau\in[0,1]$ on the validation split and choose a threshold $\tau_{\mathrm{base}}$ that
maximizes validation macro-F1. This threshold is then fixed and used on the test split, giving the decision rule
\begin{equation}
\hat{y}_{\mathrm{base}}(x)=
\begin{cases}
\text{critical}, & s_{\mathrm{base}}(x)\ge \tau_{\mathrm{base}},\\[2pt]
\text{normal},   & s_{\mathrm{base}}(x)<  \tau_{\mathrm{base}}.
\end{cases}
\label{eq:predict}
\end{equation}
For all models, once the binary decision $\hat{y}(x)$ is fixed at time $T=0$, it is treated as fixed; the
subsequent auditing step only explains this decision and never changes it.

\begin{figure}[t!]
    \centering
    \includegraphics[width=\columnwidth]{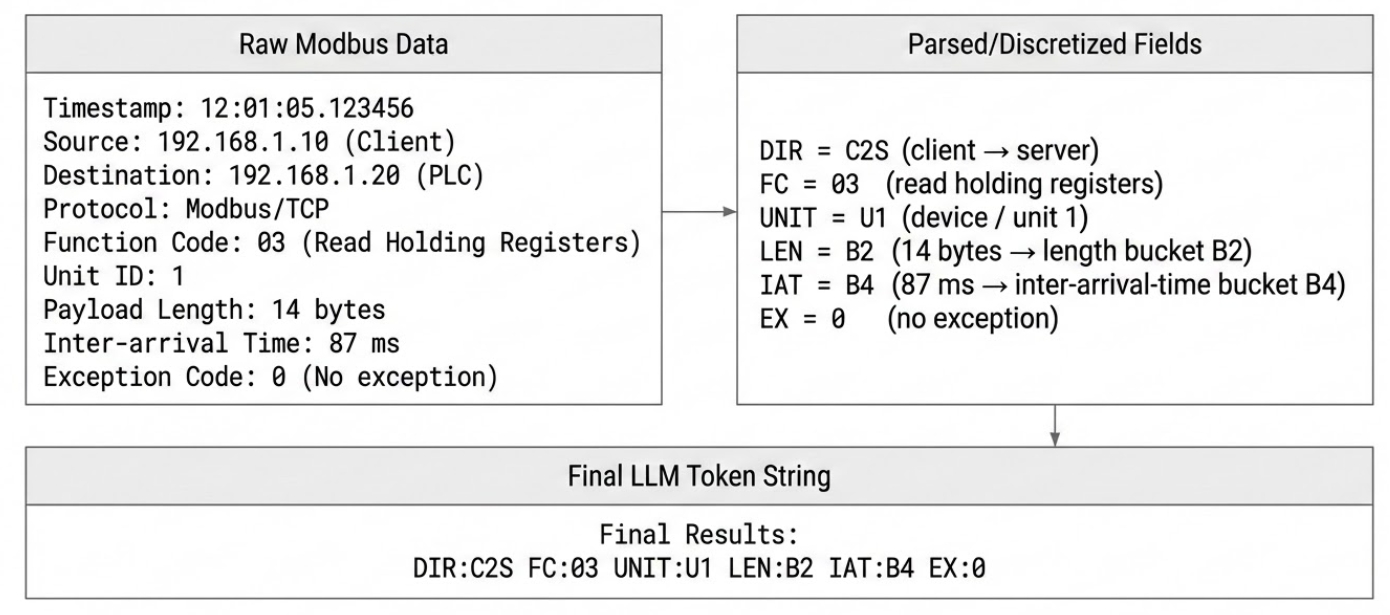}
    \caption{Example of raw-to-token encoding. A raw Modbus/TCP packet is parsed and selected fields are discretized into buckets (e.g., payload length 14 bytes $\rightarrow$ B2, inter-arrival time 87 ms $\rightarrow$ B4) to form the compact token string used by the classifier. The IP addresses shown are illustrative; the encoding retains only traffic direction (DIR) rather than raw addresses.}

    \label{fig:token_example}
\end{figure}

\subsection{Input representation}
Each Modbus frame is mapped to a short token string that records only a few protocol and timing
features that we later expose to the LLM. We encode the direction (client-to-server or server-to-client),
function code, a unit-identifier bucket, a payload-length bucket, an inter-arrival-time bucket, and a small
exception flag. Continuous quantities such as payload length and inter-arrival time are discretized into a
small number of buckets using boundaries fitted on the training split and reused unchanged on validation and
test to avoid leakage.
For example, a typical Modbus frame is encoded as
\begin{quote}
\small\mbox{\texttt{DIR:C2S FC:03 UNIT:U1 LEN:B2 IAT:B4 EX:0}}
\end{quote}
meaning  a client-to-server Modbus request with function code~3 to unit~1, with a medium-length payload, arriving
after a relatively long silence and with no Modbus exception.
The same token schema (field names and bucket identifiers) is used for both the LeMay and CIC datasets. The only
dataset-specific changes occur in the domain prompt configuration that highlights generic safety cues (e.g., certain function codes, exception flags, or timing buckets) as potentially higher risk; model parameters and decoding settings (fixed temperature, top-p, and seed) are kept identical across datasets.
This simple textual representation keeps each sample short enough for the LLM context window while allowing the
auditor to quote individual tokens verbatim as evidence in the explanation record. An example of raw-to-token encoding is depicted in Fig. \ref{fig:token_example}.

\subsection{Prompt-Configured Classifier}

Modern instruction-following LLMs can perform a wide range of classification and reasoning tasks
from natural-language instructions and a few examples \cite{Brown2020GPT3,Liu2023PromptSurvey}.
In our setting, the Modbus classifier in \eqref{eq:clf} is instantiated by a fixed LLM together
with a compact, domain-specific prompt configuration $\pi_{\text{cls}}$. Given the token string $x$
and $\pi_{\text{cls}}$, the prompted language model defines a score for each label
$y \in \{\text{normal},\text{critical}\}$, and we predict
\begin{equation}
\hat{y}(x) = \operatorname*{arg\,max}_{y \in \{\text{normal},\text{critical}\}} F(x,y;\pi_{\text{cls}}),
\end{equation}
where $F$ is determined by the fixed LLM and the classifier prompt configuration.
The configuration $\pi_{\text{cls}}$ consists of: (i) a system message describing the role of an OT
security analyst, (ii) an explanation of the token schema (DIR, FC, UNIT, LEN, IAT, EX), and
(iii) a small set of prototypical normal and critical examples together with high-level ICS safety
principles. The prompt asks the model to reason about Modbus function-code semantics, direction
and timing patterns, and their potential physical impact, and to compare the current token pattern
to the prototypes in terms of how many small token edits would be required to match them.
This ``prototype and edit-distance'' style guidance is encoded entirely in natural language and does
not rely on any learned parameters beyond the base LLM.
At implementation time, $f_{\text{cls}}(x;\pi_{\text{cls}})$ is realized as a single call to the
LLM that is required to return a top-level JSON object of the form:
\begin{lstlisting}
{ "label": "<string>",
  "confidence": <number>,
  "rationale": "<string>" }
\end{lstlisting}
The field \texttt{"label"} is constrained to either \texttt{"normal"} or \texttt{"critical"}.
The field \texttt{"confidence"} serializes $s(x)$ and is treated as an uncalibrated score recorded
for audits; it is not calibrated or re-thresholded when choosing the label. The \texttt{"rationale"}
field is a short, one-sentence natural-language justification that is stored for operator inspection
but does not affect the classifier’s decision. Deterministic decoding (fixed temperature, top-$p$
and seed) ensures that identical inputs yield identical classifier outputs, and operating thresholds
selected on the validation set are frozen for the test set.

The classifier prompt encodes a safety-first decision policy: tokens that indicate generic protocol
anomalies (e.g., exception flags, rare function codes, or unusually short or long inter-arrival times)
bias the model toward the critical label, while tokens that reflect routine background traffic
(keep-alive exchanges, typical request--response pairs) support the normal label but are not intended
to overturn clearly suspicious activity on their own. These safety cues refer only to generic token
types (such as function-code ranges and timing buckets) rather than packet-specific signatures or
dataset identifiers.

The structured explanation record used in our audits---evidence spans $E(x)$, risk tags $R(x)$, and
a counterfactual edit $c(x)$---is not produced by this first pass. Instead, it is constructed in a
separate second pass by an auditor $g$ that receives $x$ together with
the fixed decision $\hat{y}(x)$. The classifier and auditor thus share the same underlying model but
use different prompts, and the auditor is not permitted to change the classifier’s label.

\subsection{Auditor Design and Evidence Constraints}

In the second pass, given the token string $x$ and the fixed binary decision $\hat{y}(x)$,
we call an auditor $g$ that constructs a structured explanation record while keeping
$\hat{y}(x)$ unchanged. The auditor shares the same underlying LLM as the classifier
but uses a separate prompt configuration $\pi_{\text{xai}}$ and is not permitted to
change the label.

Given $\hat{y}(x)$, the auditor produces
\begin{equation}
g(x,\hat{y}) = \big(E(x),\, R(x),\, c(x)\big).
\end{equation}
The evidence list $E(x)=\{e_1,\ldots,e_m\}$ contains short substrings copied verbatim
from the input $x$ in their original order. The tag set $R(x)$ summarizes which
\emph{risk categories} support the decision (for example, function-code,
timing-related, or exception-related cues) without exposing implementation tables.
The counterfactual $c(x)$ is a single ``what-if'' token edit written as
\texttt{TOKEN\_a $\rightarrow$ TOKEN\_b} that references an existing token in $x$
and is intended to flip the label if applied. The auditor never alters $\hat{y}(x)$;
its role is to justify the decision and to propose a minimal diagnostic edit.

Faithfulness is established by intervention on the classifier, not by description
alone: we test \emph{sufficiency} by preserving only the cited evidence tokens and
\emph{necessity} by removing these tokens and keeping the rest. Because $s(\cdot)$
is an uncalibrated confidence score, $\Delta s$ is treated as a directional
diagnostic rather than a calibrated probability shift; strict label flips remain
primary, with pass-rate–versus–$\varepsilon$ curves and signed confidence deltas
(per class) reported for completeness.

\paragraph{Sufficiency}
Construct a minimal input $x_E$ that concatenates only the cited evidence in order
and rescore with the same classifier:
\begin{equation}
\big(\hat{y}_{\text{keep}},\, s_{\text{keep}}\big)
= f_{\text{cls}}(x_E;\pi_{\text{cls}}).
\end{equation}
Sufficiency passes if $\hat{y}_{\text{keep}} = \hat{y}(x)$ and
$s_{\text{keep}} \ge s(x) - \varepsilon$, with a nonnegative tolerance
$\varepsilon$ controlling allowable confidence decrease.

\paragraph{Necessity}
Remove all cited evidence to form $x_{\setminus E}$ and rescore:
\begin{equation}
\big(\hat{y}_{\text{drop}},\, s_{\text{drop}}\big)
= f_{\text{cls}}(x_{\setminus E};\pi_{\text{cls}}).
\end{equation}
We report the flip indicator $\mathbf{1}\{\hat{y}_{\mathrm{drop}} \neq \hat{y}(x)\}$
and the decrease $\Delta^{-}= s(x)-s_{\mathrm{drop}}$ with $\Delta^{-}\ge 0$;
larger values indicate greater weakening. At tolerance $\varepsilon$, a case passes
if $\hat{y}_{\mathrm{drop}} \neq \hat{y}(x)$ or $\Delta^{-}\ge \varepsilon$.

\paragraph{Counterfactual efficacy}
Apply the single edit $c(x)$ to obtain $x'$ and rescore:
\begin{equation}
\big(\hat{y}_{\text{cf}},\, s_{\text{cf}}\big)
= f_{\text{cls}}(x';\pi_{\text{cls}}).
\end{equation}
We measure label flips $\mathbf{1}\!\left[\hat{y}_{\text{cf}} \ne \hat{y}(x)\right]$
and directional change $s_{\text{cf}} - s(x)$ toward the requested label.
Counterfactuals are used as a diagnostic of boundary sensitivity rather than a
primary optimization objective. When needed, an independent checker can be used
to validate flips externally.

\subsection{Baselines}
All baselines are evaluated on the same held-out test set as the LLM. To ensure input parity, the DistilBERT baseline consumes exactly the same compact token string representation as the LLM classifier, with only minimal sanitation in the loader. The remaining supervised baselines operate on a numeric view deterministically parsed from the same ingestion pipeline, using protocol-aware features derived from function code, length bin, inter-arrival-time bin, exception code, and direction. No external signals beyond what the LLM sees are introduced.
The model family covers logistic regression, gradient-boosted trees, Kolmogorov–Arnold Networks, and a compact transformer encoder (DistilBERT). For the tabular learners we apply standardization where appropriate and enable class balancing; where supported, we employ validation-based early stopping and restore the best checkpoint. For the transformer we augment the tokenizer with domain tokens to avoid subword fragmentation on protocol vocabulary and train with a class-weighted loss. We intentionally avoid heavy hyper-parameter search so that comparisons reflect representation and learner class rather than tuning budget.
To reduce leakage from near-duplicate flows, data are split using stratification or, when identifiers permit, group-aware partitions; the resulting train/validation/test partitions are then shared across methods. We report accuracy and macro-F1 on the common test set and export per-example predictions for downstream error analysis. These choices yield reasonably strong baselines while keeping the protocol simple and consistent with the LLM setting.

\subsection{Evaluation}
All methods share a fixed train, validation, and test split drawn once. Discretization boundaries for continuous fields are estimated on the training split and reused on validation and test. For supervised baselines, the threshold \(\tau\) is selected on the validation set to maximize macro-F1 and then held fixed on the test set. The language model and all baselines are evaluated on the same held-out test split (no sub-sampling). We report accuracy, macro-F1, recall on the critical class, and F1 on the critical class. Latency and cost are summarized using medians and percentiles, and rare long-tail retries are treated as outliers. Decoding is deterministic and data-processing seeds are fixed to ensure reproducibility.

\tikzset{
  >=Latex,
  tblock/.style={
    draw, rounded corners, very thick,
    align=center,
    inner sep=2pt,
    minimum width=32mm,
    minimum height=7mm
  },
  smallblock/.style={
    draw, rounded corners, very thick,
    align=center,
    inner sep=2pt,
    minimum width=28mm,
    minimum height=6mm,
    font=\scriptsize
  },
  solidarrow/.style={->, very thick},
  note/.style={font=\scriptsize}
}

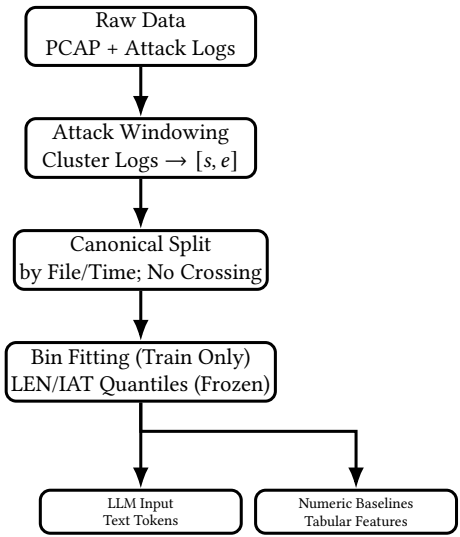
\begin{figure}[t!]
\centering
\resizebox{0.72\columnwidth}{!}{
\begin{tikzpicture}[node distance=7mm]

\node[tblock] (raw) {Raw Data\\PCAP + Attack Logs};

\node[tblock, below=of raw] (wins)
    {Attack Windowing\\Cluster Logs $\to [s,e]$};
\draw[solidarrow] (raw) -- (wins);

\node[tblock, below=of wins] (split)
    {Canonical Split\\by File/Time; No Crossing};
\draw[solidarrow] (wins) -- (split);

\node[tblock, below=of split] (bins)
    {Bin Fitting (Train Only)\\LEN/IAT Quantiles (Frozen)};
\draw[solidarrow] (split) -- (bins);

\node[smallblock, below=12mm of bins] (btext)
    {LLM Input\\Text Tokens};
\node[smallblock, below=12mm of bins, xshift=30mm] (num)
    {Numeric Baselines\\Tabular Features};

\draw[solidarrow] (bins.south) -- ++(0,-4mm) -| (btext.north);
\draw[solidarrow] (bins.south) -- ++(0,-4mm) -| (num.north);

\end{tikzpicture}}
\caption{Pipeline and leakage guards. Preprocessing includes UTC
normalization, aggregation of attacker logs into contiguous time windows,
and a canonical train/validation/test split that prevents temporal leakage
across windows. LEN/IAT bins are fit once on the training portion and then
frozen for validation and test, and each flow is branched into a compact
LLM token string and a numeric feature vector.}

\label{fig:data-timeline-vertical-skinny}
\end{figure}

\section{Experimental ICS Datasets and Preprocessing}
\label{sec:dataset}

We study network-side intrusion detection on Modbus traffic as a binary task with the positive class labeled critical. Two public captures are used. LeMay provides labeled Modbus traces with both normal and attack activity and is widely adopted in SCADA research. CIC offers a recent Modbus capture with packet-level pcaps and attack logs; attacks include reconnaissance, query flooding, false-data injection, length manipulation, stacked frames, brute-force writes, delayed responses, and replay \cite{LeMay2016SCADA,CICModbus2023}. All approaches use the same labels and a fixed train/validation/test split for fair comparison. 
Fig. \ref{fig:data-timeline-vertical-skinny}
summarizes the shared preprocessing pipeline and the leakage-control safeguards used to ensure fair comparison across train, validation, and test splits.
Labels in CIC are derived by aligning attack logs with packet timestamps on a common time base (no additional timezone adjustment), forming time windows and labeling any frame within a merged window as critical. This yields binary labels consistent with the attack schedule while abstracting away low-level choices such as the exact window gap and tail durations.
Raw records are deduplicated and time-aligned, malformed entries are removed, and remaining fields are normalized to a compact set of protocol and timing evidence sufficient for downstream modeling. To prevent leakage, discretization boundaries for continuous quantities are estimated on the training split and reused unchanged on validation and test.
To make the textual and numeric views derived from the same parse, the ingestion pipeline renders each packet into the same compact token string that is passed to the LLM, while, in the same pass, emitting tabular fields from the identical parse. Function codes are rendered as \texttt{FC:xx}, directions as \texttt{DIR:C2S/S2C}, unit identifiers as \texttt{UNIT:U\#}, and exception codes as \texttt{EX:\#}. Continuous attributes are discretized on the training split only and then frozen: payload length is mapped to \texttt{LEN:B1--B4} and inter-arrival time to \texttt{IAT:B0--B4}, with ties resolved toward the lower bin and extreme values clipped. The numeric view reuses exactly these fields. Parsing preserves packet order, is robust to missing fields via explicit NA tokens, and is fully deterministic with fixed seeds and versioned configuration, so any performance differences arise from modeling rather than preprocessing.
From the shared ingestion pipeline we derive two coordinated views of the same data. A compact textual view is consumed by the language-model classifier and allows the auditor to quote short evidence spans verbatim from the input. A numeric view is consumed by supervised baselines drawn from standard models. Both views originate from the same labels and split, ensuring head-to-head comparability; evaluation protocol details are provided in Section \ref{sec:methodology}.

\section{Results}
\label{sec:results}

Per the evaluation protocol defined in Methods, all approaches use the shared split and a single held-out test set under identical preprocessing and threshold selection. We report accuracy, macro-F1, and recall/F1 for the \textit{critical} class. Results are grouped by dataset  CIC (Table~\ref{tab:cic-acc}) and LeMay (Table~\ref{tab:LeMay-acc})—to enable head-to-head comparison; latency and cost are summarized by medians and percentiles.

\subsection{Predictive performance}

\subsubsection{CIC Dataset}
Table~\ref{tab:cic-acc} reports CIC results. Baselines cluster around accuracy near 0.97 with recall on the critical class near 0.88. On the same held-out test split, the LLM attains accuracy near 0.984 with recall on the critical class near 0.963. For baselines, errors are predominantly false negatives, consistent with the observed recall levels.
On the common test set, LLM errors are dominated by a small number of false positives where strong risk evidence co-occurs with otherwise benign context; false negatives are rare. Classical baselines show the opposite tendency on a subset of borderline cases. This complementarity suggests a practical ensemble/triage use: prioritize LLM recall for safety and surface its flagged evidence to operators, while numeric baselines screen out obvious benign traffic at scale.

\subsubsection{LeMay Dataset}
Table~\ref{tab:LeMay-acc} summarizes LeMay dataset results. The LLM reaches high recall with accuracy around 0.98 on the same held-out test split, while classical baselines are near-saturated on that split. Because all methods are evaluated on the same test set, per-cell comparisons are directly interpretable. Qualitatively, LLM errors on this set are dominated by a small number of false positives, consistent with a recall-leaning policy.

\begin{table}[t!]
\centering
\caption{Test performance on the CIC dataset}
\label{tab:cic-acc}
\resizebox{\columnwidth}{!}{%
\begin{tabular}{lcccc}
\toprule
Method & Accuracy & Macro-F1 & Recall (critical) & F1 (critical) \\
\midrule
LightGBM           & 0.9756 & 0.9658 & 0.8824 & 0.9341 \\
Logistic Regression& 0.9669 & 0.9531 & 0.8612 & 0.9207 \\
KAN                & 0.9718 & 0.9595 & 0.8729 & 0.9288 \\
DistilBERT         & 0.9739 & 0.9627 & 0.8791 & 0.9326 \\
LLM (prompt-only)  & 0.9842 & 0.9829 & 0.9625 & 0.9804 \\
\bottomrule
\end{tabular}%
}
\end{table}

\begin{table}[t!]
\centering
\caption{Test performance on the LeMay dataset}
\label{tab:LeMay-acc}
\resizebox{\columnwidth}{!}{%
\begin{tabular}{lcccc}
\toprule
Method & Accuracy & Macro-F1 & Recall (critical) & F1 (critical) \\
\midrule
LightGBM            & 0.9993 & 0.9993 & 1.0000 & 0.9994 \\
Logistic Regression & 0.9965 & 0.9964 & 1.0000 & 0.9969 \\
KAN                 & 0.9972 & 0.9971 & 0.9966 & 0.9975 \\
DistilBERT          & 0.9965 & 0.9964 & 0.9950 & 0.9968 \\
LLM (prompt-only)   & 0.9822 & 0.9820 & 1.0000 & 0.9838 \\
\bottomrule
\end{tabular}%
}
\end{table}

\subsection{Explanation audits}

\begin{figure}[t!]
  \centering
  \includegraphics[width=\linewidth]{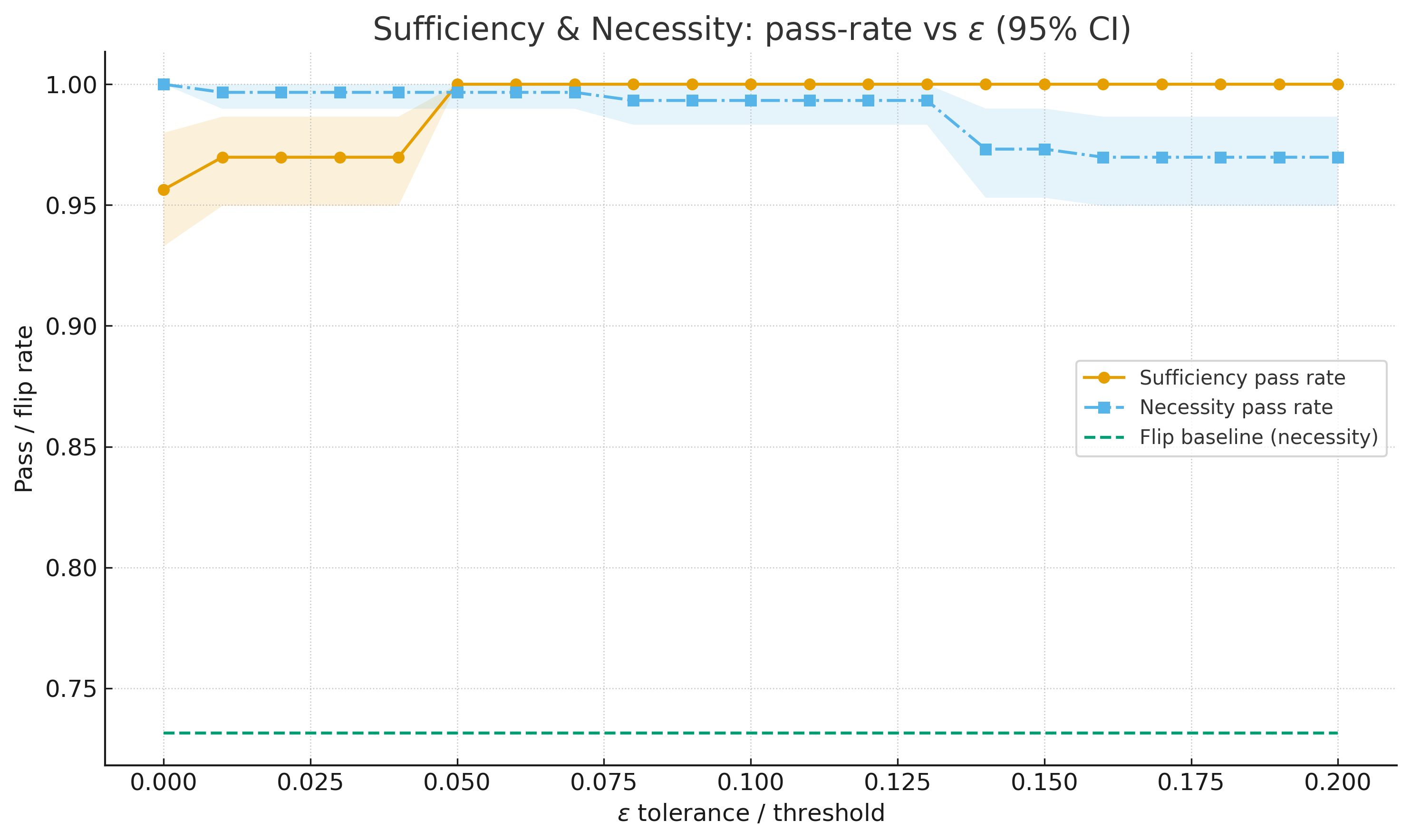}
  \caption{Pass-rate vs $\varepsilon$ for sufficiency and comprehensiveness (95\% CIs); dashed line shows the label flipped baseline.}
  \label{fig:sn-pass-eps}
\end{figure}

Explanations are structurally valid and directly tied to the input tokens. JSON well-formedness is 100\% and span validity (verbatim evidence grounding) is 100\%: This is a syntactic grounding check, every evidence item is a literal substring of the corresponding token string $x$. For example, if the input contains tokens such as \texttt{FC:129} and \texttt{IAT:B3} and the classifier assigns a critical label, the auditor is required to copy these exact tokens into the evidence list and to attach the corresponding risk tags (e.g., function-code and timing rules). In our audits we do not observe cases where such designated high-risk tokens are omitted from the explanation. When any high-risk tokens occur, they are consistently cited in the evidence and mirrored in the risk tags (100\%). We also do not find instances where the explanation points in the opposite direction of the label (for example, a \emph{normal} label justified only by high-risk tags). Among critical predictions, 1.13\% of cases have no cited high-risk tokens at all; these rare “unexplained critical’’ items are flagged for silent
retry and monitoring.
For Sufficiency, keeping only the cited evidence tokens $E(x)$ (the tokens that the auditor copied verbatim from the
input) preserves the original decision in 95.64\% of cases under a no-decrease criterion, with score
changes concentrated near zero. By class, the mean change is +0.0018 for critical and $-0.0035$ for
normal, with ranges approximately $[-0.01, 0.04]$ and $[-0.05, 0]$, respectively. Under stricter
requirements, the pass rate is 7.72\% for $\Delta s \ge 0.01$ and 1.68\% for $\Delta s \ge 0.02$.
For Necessity, removing the cited evidence tokens $E(x)$ substantially weakens decisions: the overall flip rate is
73.15\% with an average score decrease of 0.162 (quartiles about 0.13 / 0.25 / 0.38). By class,
critical flips occur in 95.27\% of cases with mean decrease 0.18, while normal flips occur in 44.19\%
of cases with mean decrease 0.1395. The effect scales with the number of cited tokens removed: the
average decrease rises from about 0.084 (one token) to 0.148 (two tokens) and 0.200 (three tokens).
Following ERASER~\cite{DeYoung2020ERASER}, we assess explanation faithfulness using sufficiency and comprehensiveness (necessity here). Fig.~\ref{fig:sn-pass-eps} reports pass rates across $\varepsilon$ with 95\% bootstrap confidence intervals; the dashed line shows the strict label-flip baseline when cited evidence $E$ is removed. The results indicate that the cited evidence is often close to sufficient and decision-relevant.
For each audited item, the auditor proposes a single token-level edit $c(x)$ that changes one
evidence token from the input (for example, replacing \texttt{FC:129} with \texttt{FC:003}) toward
the opposite class. We apply the edit to obtain $x'=\mathrm{apply}(x,c)$ and recompute the decision
with the same classifier and decoding/threshold settings as in the main evaluation. Success is
defined strictly as a label flip, $\hat y(x') \neq \hat y(x)$; empty edits are never counted and
invalid edits are treated as failures. On the same held-out test set, this strict single-edit flip
rate is 23.43\%.
Consistent with recent studies reporting that counterfactual self-explanations and executable counterfactuals remain challenging for current LLMs, our one-step probe yields limited directional gains and only moderate flip rates; we therefore treat it as a diagnostic of boundary sensitivity rather than a primary objective.

\subsection{Runtime latency and cost}
We evaluate end-to-end latency and amortized API cost of the LLM+XAI pipeline. To mitigate occasional network spikes, latencies are summarized after an interquartile-range (IQR) filter with Tukey fences [0.342 s, 1.585 s]. We also report the 95th/99th percentile (p95/p99) latency. With GPT-4o (text), the median latency is 0.91 s, p95 is 1.35 s, and p99 is 1.51 s. Using provider list prices for GPT-4o (\$2.50 per million input tokens, \$10.00 per million output tokens), the amortized cost is \$0.42--\$1.42 per 1k inferences depending on the assumed prompt overhead (0--400 additional tokens per call).

\section{Discussion}
Our binary normal/critical formulation deliberately simplifies the detection problem and likely makes the task easier than operational deployments. In practice, ICS monitoring often requires finer-grained severity tiers and distinct attack categories, where class boundaries may be substantially harder—especially when different events share similar Modbus token patterns. Extending the present prompt-based setup to multi-class or multi-level ICS taxonomies is therefore an important next step, though performance would likely fall relative to the binary results in Tables 1 and 2.
More broadly, the current pipeline relies primarily on prompt-based steering without external grounding beyond the prompt text. Future work should test retrieval-grounded prompting, robustness under distribution shift and operational load, and extension to real alerting workflows and additional ICS protocols.

\section{Conclusion}
\label{sec:conclusion}
We studied the feasibility of using an off-the-shelf, prompt-configured LLM as a binary normal/critical classifier for ICS/SCADA Modbus traffic on two public datasets. Under a shared evaluation protocol, the LLM achieved high detection performance broadly comparable to strong supervised baselines trained on matched inputs, while also producing a concise, token-grounded audit record for each decision. Intervention-based audits further suggest that the cited evidence tokens are often decision-relevant to the model’s own prediction. Taken together, these results support the use of prompt-configured LLMs as a transparent triage component for ICS monitoring under human review.

\bibliographystyle{ACM-Reference-Format}
\bibliography{references_pesgm_min}



\section*{Appendix: Example Prompts}

This appendix lists one representative configuration of the prompts used in our experiments.
Both public Modbus datasets use the same prompt structure, with only minor changes to token
names and cue lists.

\subsection*{Classifier prompt (binary decision)}

The prompt below shows the system message and few-shot examples used to obtain a
normal/critical decision in JSON form.

\begin{lstlisting}
### SYSTEM ROLE
You are a Senior Operational Technology (OT) Security Analyst responsible for protecting a high-voltage power grid. 
Your goal is to detect cyberattacks by analyzing Modbus/TCP traffic logs. You prioritize "Physical Safety" and "Process Integrity" above all else.

### COGNITIVE FRAMEWORK
Do NOT rely on a fixed list of "bad" function codes. Instead, apply the following FIRST PRINCIPLES to judge the risk:

1. **The Principle of Impact:** - Ask yourself: "Does this command merely *observe* the system (Read), or does it attempt to *change* the system (Write/Control)?"
   - **READ operations** (Polling) are the baseline of industrial automation and are generally **Normal**, unless they occur at an impossible speed (Flooding).
   - **WRITE/CONTROL operations** actively alter physical devices (coils, registers). In a stable grid, these are rare. Without clear context, treat unsolicited state changes as **Critical**.

2. **The Principle of Rhythm (Time Analysis):**
   - Machines are rhythmic; Hackers are bursty.
   - **Normal:** Periodic, steady inter-arrival times (e.g., regular polling intervals).
   - **Critical:** Sudden deviations from the rhythm. Extremely short intervals (B0/B1) suggest automated flooding or fuzzing. Extremely long gaps followed by activity may suggest a "low-and-slow" attack.

3. **The Principle of Protocol Compliance:**
   - Any proprietary, undefined, or malformed function codes are immediately **Critical**.
   - Exception codes (EX) usually indicate a device failure or a scanner probing invalid addresses.

### INPUT FORMAT LEGEND
You will receive a single log line with discretized tokens:
- **DIR:** Direction (C2S = Client to Server / S2C = Server to Client).
- **FC:** Modbus Function Code (Integer). Use your knowledge of the Modbus spec to interpret this.
- **UNIT:** Unit ID (e.g., U1).
- **LEN:** Payload Length Bucket (B0=Empty/Tiny ... B4=Large).
- **IAT:** Inter-Arrival Time Bucket (B0=Immediate/Burst ... B4=Long Silence).
- **EX:** Exception Code (0=None, others=Error).

### REASONING STEPS
Before deciding, perform this internal monologue:
1. **Semantic Decode:** What is the specific purpose of this FC according to the Modbus protocol standard? (e.g., Is it reading inputs or forcing coils?)
2. **Intent Check:** Does this combination of Direction + FC + Frequency look like a SCADA master polling a sensor (Benign), or an external actor trying to manipulate the grid (Malicious)?
3. **Risk Assessment:** If this command succeeds, could it physically trip a breaker or alter a sensor reading?

### OUTPUT FORMAT
Output ONLY a JSON object. 
- "label": Must be "normal" or "critical".
- "confidence": Float between 0.0 and 1.0.
- "rationale": A concise, 1-sentence explanation focusing on the *operational impact* (e.g., "Unauthorized attempt to alter device state via Write command").

\end{lstlisting}



\end{document}